\documentclass[twocolumn,showpacs,preprintnumbers,amsmath,amssymb]{revtex4}
\usepackage{graphicx}

\begin{document}

\title{Temporal shape manipulation of adiabatons}

\author{V.G.Arkhipkin}
\affiliation{L.V.Kirensky Institute of Physics SB RAS, 660036 Russia Krasnoyarsk}
\email{avg@iph.krasn.ru}
\author{I.V.Timofeev}
\affiliation{L.V.Kirensky Institute of Physics SB RAS, 660036 Russia Krasnoyarsk}
\date{\today}

\begin{abstract}We describe how to control the temporal shape of adiabaton using peculiarities of propagation dynamics under coherent population trapping. Temporal compression is demonstrated as a special case of pulse shaping. The general case of unequal oscillator strengths of two optical transitions in atom is considered. \end{abstract}


\maketitle

\section{Introduction}

Electromagnetically induced transparency (EIT) and coherent population trapping (CPT) can facilitate coherent control of light under propagation through a medium \cite{1,2}. In addition to their fundamental interest, investigations of these processes are stimulated by practical possibilities, such as manipulating a group velocity of light and light storage in atomic medium \cite{3,4}, enhanced nonlinear optical processes \cite{5}, quantum memory \cite{4} and so on. 

The CPT is a quantum interference effect and takes place under resonance interaction of two laser fields (probe and coupling) with three-level atomic systems. The essence of this effect is that under certain conditions atoms are trapped into the coherent superposition of two lower states $\vert 1 \rangle$ and $\vert 2 \rangle$, which is called CPT-state \cite{6,7}. Under CPT condition the medium becomes coherent and possesses unusual properties, many of which contradict with the intuitive views. The CPT leads to the maximal coherence at the Raman transition and the medium becomes transparent for the probe and coupling pulses \cite{5,8}. This phenomenon allows recording, storing and reading of information about strong optical pulses \cite{9,10}, to control the degree of excitation of spatially localized regions inside an absorbing three-level medium \cite{11} and to generate matched pulses \cite{12,13}, adiabatons \cite{14,15} and dressed-field pulses \cite{16}. 

Recently it was shown how EIT can be used for coherent control of the weak pulse shape \cite{17}. The idea is following. Under EIT the weak probe pulse propagates with a slow group velocity depending on an intensity of the coupling field. If the intensity of the coupling field depends on a time, different points of the probe pulse experience different values of intensity of coupling field and travel with different propagation velocities, giving rise to temporal reshaping of the probe. A proper choice of the temporal shape of the coupling pulse allows control and manipulation of the probe pulse envelope. In this paper we generalize this method for controlling the temporal shape of the intense probe pulse using the peculiarities of CPT propagation dynamics. Temporal compression of adiabatons is demonstrated as special case of pulse tailoring. 

\begin{figure}
  \includegraphics[width = 4 cm]{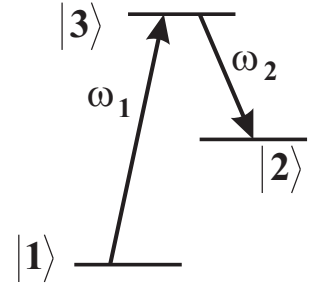}\\
  \caption{The three-level system coupled by two resonant pulses with Rabi frequencies $G_{c}$ and $G_{p}$.
$\omega_{c}$ and $\omega_{p}$ are the frequencies of the control and probe
pulses. The transition $\vert 1 \rangle - \vert 2 \rangle$ is dipole forbidden.}\label{Scheme3L}
\end{figure}

\section{Principal equations}

Consider the propagation of two laser pulses in a medium consisting of three-level atoms (Fig.~1). Pulses propagate along an axis $z$ in one direction. A probe pulse (with the slowly varying envelope $E_p (t)$ and frequency $\omega_{p}$) is tuned on resonance with $\vert 3 \rangle - \vert 1 \rangle$ transition, and the coupling pulse ($E_c (t)$, $\omega_{c})$ is tuned so that exact two-photon resonance between states $\vert 1 \rangle-\vert 2 \rangle$ is achieved. The coupling pulse is switched on earlier and switched off later than probe. The pulse durations $T_p$ and $T_c$ are much less than any relaxation times of atoms and $T_p<T_c$. Intensities of both pulses are comparable.

A propagation of pulses can be described by Schr\"{o}dinger equation and reduced wave equations for Rabi frequencies (Maxwell-Schr\"{o}dinger equations) which should be solved self-consistently. For the case when the fields are in resonance with their respective transitions, Maxwell-Schr\"{o}dinger equations are
\begin{equation}\label{eq1}
\frac{\partial }{\partial \tau }\left({{
\begin{array}{*{20}c}
 {a_1 } \hfill \\
 {a_2 } \hfill \\
 {a_3 } \hfill \\
\end{array}
}} \right) = i\left( {{
\begin{array}{*{20}c}
 0 \hfill & 0 \hfill & {G_p^\ast } \hfill \\
 0 \hfill & 0 \hfill & {G_c^\ast } \hfill \\
 {G_p } \hfill & {G_c } \hfill & 0 \hfill \\
\end{array}
}} \right)\left( {{
\begin{array}{*{20}c}
 {a_1 } \hfill \\
 {a_2 } \hfill \\
 {a_3 } \hfill \\
\end{array} }} \right),
\end{equation}

\begin{equation}\label{eq2}
\frac{\partial }{\partial \zeta }\left( {{
\begin{array}{*{20}c}
 {G_p } \hfill \\
 {G_c } \hfill \\
\end{array} }} \right) = i\left( {{
\begin{array}{*{20}c}
 {K_p a_1^\ast a_3 } \hfill \\
 {K_c a_2^\ast a_3 } \hfill \\
\end{array} }} \right).
\end{equation}
Here $\zeta = z,\;\tau = t - z/c$  --  space and time coordinates in a frame moving with light velocity $c$ in empty space; $a_{1,2,3}$ -- the probability amplitudes of atomic states; $2G_{p,c} = {E_{p,c} d_{1,2} } /\hbar $ -- the Rabi frequencies of fields; $E_{p,c}$ -- the probe and coupling field strengths; $d_{13,23}$ -- the electrical dipole moments of the relevant atomic transitions; $\hbar$ -- the Plank constant; $K_{p,c} = {2\pi N\omega _{p,c} \left| {d_{13,23} } \right|^2} / {\hbar c}$ -- the field-atomic medium coupling coefficients; $N$ -- the atomic concentration. Initially all atoms are in the ground state $\left| 1 \right\rangle $: $a_{1,2,3} (\tau = - \infty, \zeta) = (1;0;0)$. The solution of Eqs. (\ref{eq1}) and (\ref{eq2}) gives the complete evolution of the atom-field system.

The analytical solution of the equation system (\ref{eq1},\ref{eq2}) is possible only in adiabatic approximation \cite{8,14}. In this case $\left| {a_3} \right| \ll 1$ and ${G_p } / {G_c } = - {a_2} / {a_1}$. The condition $\left| {a_3 } \right| \ll 1$ means, that the population of intermediate state $\vert 3 \rangle$ is close to zero in the interaction of pulses with atoms. The population is trapped in a coherent superposition of states $\vert 1 \rangle$ and $\vert 2 \rangle$ -- the effect of CPT. Under CPT pulses do not interact with medium \cite{2,7}\textbf{.} It means that pulses can propagate practically without absorption.

In the adiabatic approximation Eqs. (\ref{eq1}) and (\ref{eq2}) lead to photon number conservation law
\begin{eqnarray*}
\frac{G_c^2 (\tau ,\zeta )}{K_c } + \frac{G_p^2 (\tau, \zeta )}{K_p } &=& \\
\frac{G_c^2 (\tau ,\zeta = 0)}{K_c } + \frac{G_p^2 (\tau ,\zeta = 0)}{K_p } &=& V(\tau ,\zeta = 0).
\end{eqnarray*}
The conservation law implies that any change in the probe pulse is compensated by a corresponding change in the coupling pulse and so $V$ does not depend on the space variable during the propagation. The input fields determine the temporal shape of $V$.

In the adiabatic approximation field equations (\ref{eq2}) have the form

\begin{equation}\label{eq3}
\frac{\partial \vec {G}}{\partial \zeta } = - \hat {{\rm K}}\frac{1}{\vec {G}^2}\frac{\partial \vec {G}}{\partial \tau },\quad\vec {G} = (G_c ,G_p ).
\end{equation}

Let us introduce new variable, mixing angle $\theta \left( {\tau ,\zeta } \right)$, which is determined as $\tan \theta = {G_p } / {G_c }$. The equation for $\theta \left( {\tau ,\zeta } \right)$ is \cite{8}

\begin{equation}\label{eq4}
\frac{\partial \theta }{\partial \zeta } + \frac{K^2\left( \theta \right)}{K_p G_c^2 + K_c G_p^2 }\frac{\partial \theta }{\partial \tau } = 0,
\end{equation}

\[
K\left( \theta \right) = {\left( {K_p G_c^2 + K_c G_p^2 } \right)} / {\vec {G}^2} = K_p \cos ^2\theta + K_c \sin ^2\theta.
\]
The mixing angle appears constant along characteristic curves

\[\zeta \left( {\tau ,\tau _0 } \right) = K^{ - 2}\left( {\theta \left( {\tau _0 ,\zeta = 0} \right)} \right)\int_{\tau _0 }^\tau {\left( {K_p G_c^2 + K_c G_p^2 } \right)\;d{\tau }',} \]
$\tau_0$ -- the time when given characteristic curve intersects the input of medium. $\tau_0$ is to be determined from last equation and the solution of Eq. (\ref {eq4}) can be written as:
\begin {equation*}
\theta(\tau,\zeta)=\theta(\tau_0,0).
\end {equation*}
All physical quantities can be expressed through the mixing angle $\theta \left( {\tau ,\zeta } \right)$ \cite{8}

\begin{figure}
  \includegraphics[width = 7 cm]{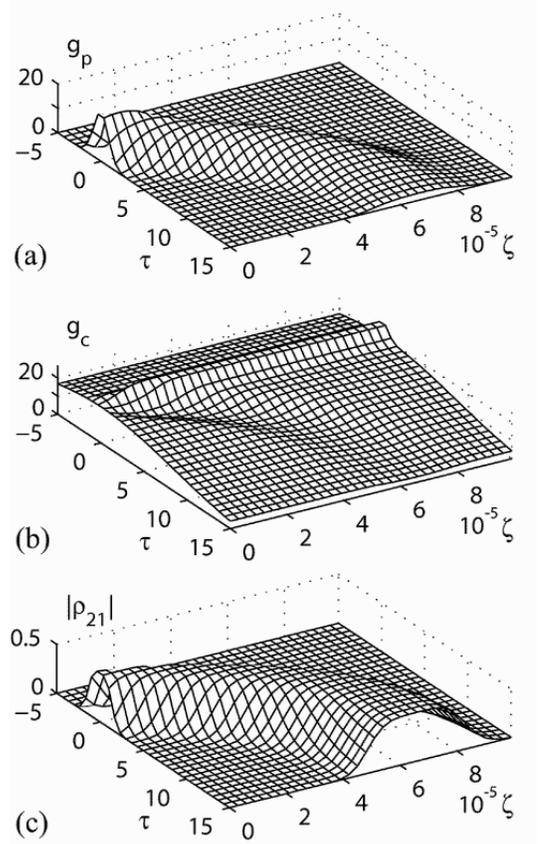}\\
  \caption{The space-time evolution of normalized Rabi frequencies of the probe $g_{p}=G_{p}T_{p}$ and control $g_{c}=G_{c}T_{p}$ pulses and atomic coherence $\rho _{21}=a_{2}^*a_{1}$. The probe and coupling pulses are the Gaussian pulses with parameters $T_{c}=10 T_{p}$ and maximal values of Rabi frequencies at the input of medium $g_{p}^{0}=g_{c}^{0}=20$. The time $\tau$ is measured in units $T_{p}$, the length $\zeta$ -- in length of linear absorption.}\label{2}
\end{figure}

\begin{equation}\label{eq5}
\left( {{\begin{array}{*{20}c} {G_p } \hfill \\ {G_c } \hfill \\\end{array} }} \right) = 2\sqrt {\frac{\left. {\left( {K_p G_c^2 + K_c G_p^2 } \right)} \right|_{\zeta = 0} }{K\left( \theta \right)}} \left( {{\begin{array}{*{20}c} {\sin \theta } \hfill \\ {\cos \theta } \hfill \\\end{array} }} \right),
\end{equation}

\begin{equation}\label{eq6}
\left(
{{\begin{array}{*{20}c}
{a_1 } \hfill \\
{a_2 } \hfill \\
{a_3 } \hfill \\
\end{array} }} \right)
 =
\left(
{{\begin{array}{*{20}c}
{\cos \theta } \hfill \\
{\sin \theta } \hfill \\
{ - \left| {{\partial (\vec {G}} / {\vec {G}^2)/\partial \tau }} \right|} \hfill \\
\end{array} }} \right).
\end{equation}
The solution (\ref{eq5}) and (\ref{eq6}) can be applied only within the area of adiabaticity which is limited by the relation \cite{14}

\[
\vert G_c \frac{\partial G_p }{\partial \tau } - G_p \frac{\partial G_c }{\partial \tau }\vert \ll (G_c^2 + G_p^2 )^{3/2}.
\]

In contrast to usual steady state solution which does not depend on initial conditions, the space-time evolution of the probe and coupling pulses under CPT conditions depends on the pulse forms at the input of medium. Some aspects of this dependance are discussed in \cite{8}.

\section{Temporal shape control \protect\\ of the probe pulse by CPT: \protect\\ compression of pulses}

Figure 2 demonstrates the evolution of the Rabi frequencies of pulses and the atomic coherence under CPT in the case of the Gaussian pulses at the input of medium, also $K_{p}=K_{c}$ and $T_{c}=10 T_{p}$. It is visible, that the probe pulse is gradually depleted and the control gets stronger. Note that the pulse shape at the initial stage of propagation shows very little variation with the length, which may considerably exceed the linear absorption length. Complete reemitting of the probe pulse into the control one during propagation is possible. The atomic or Raman coherence is excited only in a part of medium, that is spatially localized. Outside of this area, atoms remain unexcited in the ground state. The spatial distribution of atomic coherence keeps the information about pulses. This can be used for record and storage of information about the probe pulse in the CPT-modified medium \cite{9,10}.

In a case, when $T_{p} \ll T_{c}$ and the amplitude of the coupling pulse is constant, the probe and coupling pulses have complementary envelopes and propagate without form variation and with equal group velocity. Such pulses are called adiabatons \cite{14}.

Under unequal propagation constants $K_{p,c}$ (unequal oscillator strengths of two optical transitions in the atom) the adiabatons are not shape-preserving but undergo a front sharpening (Fig.~3): under $K_p < K_c$ a back edge becomes steeper (dash-dot line) , and under $K_p > K_c$ - a leading edge becomes steeper (dashed line).
\begin{figure}\label{3}
  \includegraphics[width = 7 cm]{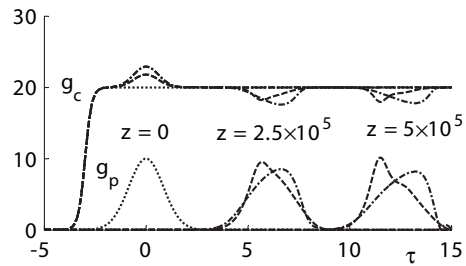}
  \caption{Propagation of the adiabatons ($g_{p}=G_{p}T_{p}$, $g_{c}=G_{c}T_{p})$ for different propagation distances within the medium in the case$K_p \ne K_c $: $K_{c }/K_{p}=1.25$ -- dash-dot line, $K_{c}/K_{p} = 0.75$ - dashed.}
\end{figure}

Since under CPT the space-time evolution of the probe pulse depends on the temporal shape of the coupling pulse, we can manipulate the shape of the probe pulse by proper choice of the coupling pulse envelope at the entry of medium. In this regard CPT can be viewed as a way of the coherent control of temporal pulse shaping. In particular, it is possible to choose such coupling pulse shape, that the trailing edge of the probe pulse travels faster than leading one. This results in the compression of probe pulse. Figure 4 demonstrates an example of the temporal compression of probe pulse using coupling pulse with the envelope shown in Fig.~4a (dashed line). A time evolution of pulses is much similar to adiabatons propagation \cite{14}. Pulse propagation in this case can be treated as adiabatonic pair extended to time shape variation (quasi-adiabatons) since both pulse envelopes vary coherently and travel with equal velocity. The reason of compression is that the leading edge of probe pulse is slowed down more strongly than the trailing one. As a result the pulse is compressed in time under propagation through the medium.

\begin{figure}
   \includegraphics[width = 7 cm]{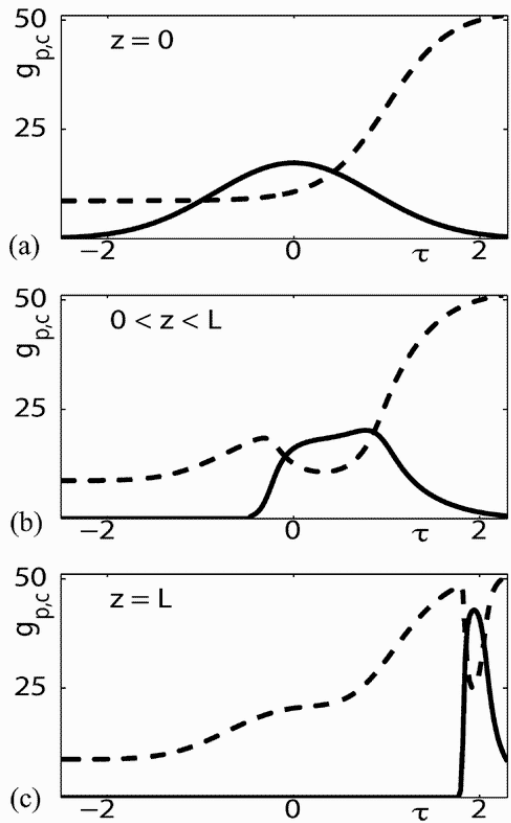}\\   
   \caption{A compression of the probe pulse in the case $K_{p}=K_{c}$. Temporal profiles of the normalized Rabi frequencies of the probe $g_{p}=G_{p}T_{p}$ and control (dashed line) $g_{c}=G_{c}T_{p}$ pulses at different propagation distances within the medium. (a) At the input of medium $z=0$; (b) at some distance within the medium; (c) at the output of medium $z=L.$}\label{}
\end{figure}

 Note that the compression effect is independent on the detailed temporal structure of the coupling pulse. The compression takes place also under a linear growth of amplitude of the coupling pulse. A finite spectral bandwidth of transparency window sets a limit to the temporal duration of the probe pulse that can travel in the medium without absorption.

The pulse compression takes place also in the case of unequal propagation constants $K_{p,c}$, which are defined by the oscillator strengths of transitions, as shown on Fig.~5.

\begin{figure}
   \includegraphics[width = 6 cm]{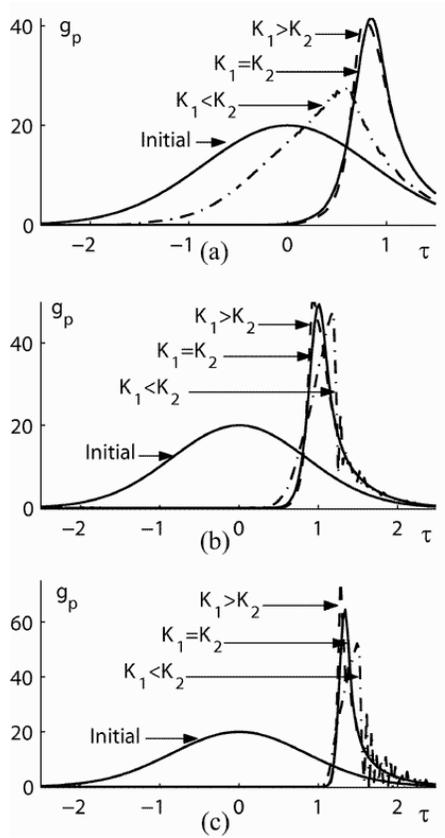}\\   
   \caption{The time evolution of Rabi frequency of probe pulse $g_{p}$ at the different depth in the medium: $z_{1}=L/9$; b) $z_{2}=L/2$; c) $z_{3}=L$. The solid line -- $K_{p}=K_{c}$, dashed - $K_{p}=4K_{c}$, dashed-dot - $4K_{p}=K_{c}$ }\label{}
\end{figure}

Pulse compression is a particular case of temporal shaping. In the general case, the proper choice of the temporal shape of the coupling pulse allows to obtain probe pulse with different temporal shapes at output. For example, we can obtain a flat-top pulse (Fig.~6) or two-peaked pulse (Fig.~7) like in \cite{17}.
\begin{figure}
   \includegraphics[width = 6 cm]{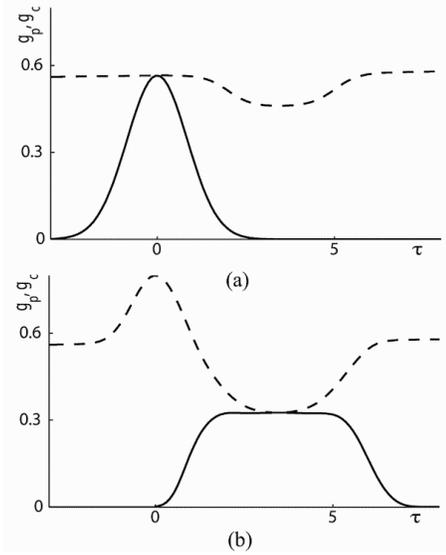}\\
   \caption{Flat-top pulse: Time evolution of normalized Rabi frequencies of probe $g_p$ (continuous curve) and coupling $g_c$ (dashed-line curve) pulses at medium inlet (a) and at medium output (b).}\label{} 
\end{figure}
For the results presented we have checked that the numerical solution of the Maxwell-Schr\"{o}dinger equations and obtained analytical expression provide exactly the same results.

\begin{figure}
   \includegraphics[width = 6 cm]{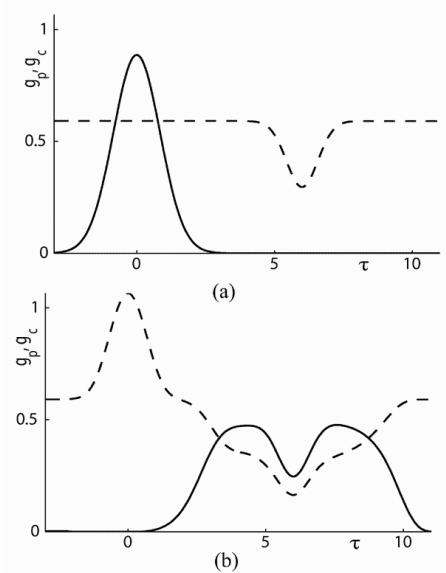}\\   
   \caption{Two-peaked pulse: Time evolution of normalized Rabi frequencies of probe  $g_p$   (continuous curve) and coupling $g_c$ (dashed-line curve) pulses at medium inlet (a) and at medium output (b)}\label{} 
\end{figure}

In conclusion, we have shown that temporal pulse compression can be achieved using CPT schemes. These processes present both fundamental interest and applications in nonlinear optics, because the compressed pulse as a light source can increase the efficiency of nonlinear processes. 

This work was supported by Russian Foundation for Basic Research (grant  02–02–16325) and Krasnoyarsk Regional Science Foundation (grant 12F0042c).


\begin{references}

\bibitem{1} S.E. Harris, Physics Today \textbf{50}, 36 (1997).

\bibitem{2} M.O. Scully, M.S. Zubairy, Quantum optics, Cambridge University, 1997.

\bibitem{3} A.B. Matsko, O. Kocharovskaya, Yu. Rostovtsev, et.al., Edv. in At., Mol. and Opt. Phys., \textbf{46}, 191 (2001); D.F. Phillips, A. Fleischhauer, A. Mair, R.L. Walsworth, M.D. Lukin, Phys.Rev.Lett. \textbf{86}, 783 (2001); Chien Liu, Z. Dutton, C.H. Behroozi et.al., Nature 409, (2001); V.G. Arkhipkin, I.V. Timofeev, JETP Lett., \textbf{76}, 66 (2002).

\bibitem{4} M.D. Lukin, Rev. Mod. Phys., \textbf{75}, 457 (2003); M.D. Lukin, A. Imamoglu, Nature, \textbf{412}, 273 (2001).

\bibitem{5} M.D. Lukin, P.R. Hemmer, M.O. Scully, Edv. in At., Mol. and Opt. Phys., \textbf{42}, 347 (2000); S.E. Harris, J.E. Field, A. Imamoglu, Phys.Rev.Lett. \textbf{64}, 1107 (1990); V.G. Arkhipkin, S.A. Myslivets, Quantum Elecronics, \textbf{25}, 901 (1995); P.R. Hemmer, M.S. Shahriar, J. Donoghue et.al., Opt.Lett. \textbf{20}, 982 (1995); S.E. Harris, M. Jain, Opt.Lett. \textbf{22}, 636 (1997); V.G. Arkhipkin, D.M. Manushkin, S.A. Myslivets, et.al. Quantum Electronics, \textbf{28}, 637 (1998); L. Deng, M.G. Payne, W.R. Garrett, Phys.Rev. A\textbf{58}, 707 (1998).

\bibitem{6} B.D Agap'ev. et.al., Uspehi Fis.Nauk, \textbf{163}, 1 (1993).

\bibitem{7} E. Arimondo, 1996, in \textit{Progress in Optics}, edited by E.Wolf (Elsevier, Science, Amsterdam), \textbf{35}, p.257.

\bibitem{8} V.G. Arkhipkin, I.V. Timofeev, Phys.Rev.A, \textbf{64}, 053811 (2001).

\bibitem{9} V.G Arkhipkin, V.P Timofeev, I.V. Timofeev, 2003, Luminescence and Laser Physics, Proc. Int.  school-seminar 23-28.09.2002, Irkutsk, Irkutsk University 2003; pp.19-26 (in Russian); Radiophysics and Quantum Electronics, \textbf{47}, 811 (2004).

\bibitem{10} T.N. Dey, G.S. Agarwal, Phys.Rev.A, \textbf{67}, 033813 (2003).

\bibitem{11} J.R. Csesznegi, B.K. Clark, R. Grobe, Phys.Rev. A\textbf{57}, 4860 (1998).

\bibitem{12} S.E. Harris, Phys.Rev.Lett., \textbf{72}, 52 (1994); J.H. Eberly, Quantum Semiclass. Opt., \textbf{7}, 373 (1995); G. Vemuri, K.V. Vasavada, G.S. Agarwal, Q. Zhang, Phys.Rev.A, \textbf{54},3394 (1996).

\bibitem{13} E. Cerboneschi, E. Arimondo, Phys.Rev. A\textbf{52}, R1823 (1995); Phys.Rev. A\textbf{54}, 5400 (1996); A. Merriam, S.J. Sharpe, D. Manuszak et.al., Phys.Rev.Lett., \textbf{84,} 5308 (2000); V.G. Arkhipkin, I.V. Timofeev, Proc. All-Russian seminar "Simulation of nonequilibrium systems - 2002", Krasnoyarsk, 2002. p.5-6 (in Russian).

\bibitem{14} R. Grobe, F.T. Hioe, J.H. Eberly, Phys.Rev.Lett., \textbf{73}, 3183 (1994); M. Fleischhauer, A.S. Manka, Phys.Rev.A, \textbf{54},794 (1996).

\bibitem{15} I.E. Mazets, Phys.Rev.A, \textbf{54}, 3539 (1996).

\bibitem{16} J.H. Eberly, M.L. Pons, H.R. Haq, Phys.Rev.Lett., \textbf{72}, 56 (1994).

\bibitem{17} R. Buffa, S. Cavalieri, M.V. Tognetti, Phys.Rev.A, \textbf{69}, 033815 (2004).

\end{references}
\end{document}